\input harvmac.tex
\input epsf.tex
\input amssym
\input ulem.sty
\input graphicx.tex


\let\includefigures=\iftrue
\let\useblackboard=\iftrue
\newfam\black

\def\figin{\epsfcheck\figin}\def\figins{\epsfcheck\figins}
\def\epsfcheck{\ifx\epsfbox\UnDeFiNeD
\message{(NO epsf.tex, FIGURES WILL BE IGNORED)}
\gdef\figin##1{\vskip2in}\gdef\figins##1{\hskip.5in}
\else\message{(FIGURES WILL BE INCLUDED)}%
\gdef\figin##1{##1}\gdef\figins##1{##1}\fi}
\def\DefWarn#1{}
\def\figinsert{\goodbreak\midinsert}
\def\ifig#1#2#3{\DefWarn#1\xdef#1{fig.~\the\figno}
\writedef{#1\leftbracket fig.\noexpand~\the\figno} %
\figinsert\figin{\centerline{#3}}\medskip\centerline{\vbox{\baselineskip12pt
\advance\hsize by -1truein\noindent\footnotefont{\bf
Fig.~\the\figno:} #2}}
\bigskip\endinsert\global\advance\figno by1}


\includefigures
\message{If you do not have epsf.tex (to include figures),}
\message{change the option at the top of the tex file.}
\input epsf
\def\figin{\epsfcheck\figin}\def\figins{\epsfcheck\figins}
\def\epsfcheck{\ifx\epsfbox\UnDeFiNeD
\message{(NO epsf.tex, FIGURES WILL BE IGNORED)}
\gdef\figin##1{\vskip2in}\gdef\figins##1{\hskip.5in}
\else\message{(FIGURES WILL BE INCLUDED)}%
\gdef\figin##1{##1}\gdef\figins##1{##1}\fi}
\def\DefWarn#1{}
\def\figinsert{\goodbreak\midinsert}
\def\ifig#1#2#3{\DefWarn#1\xdef#1{fig.~\the\figno}
\writedef{#1\leftbracket fig.\noexpand~\the\figno}%
\figinsert\figin{\centerline{#3}}\medskip\centerline{\vbox{
\baselineskip12pt\advance\hsize by -1truein
\noindent\footnotefont{\bf Fig.~\the\figno:} #2}}
\endinsert\global\advance\figno by1}
\else
\def\ifig#1#2#3{\xdef#1{fig.~\the\figno}
\writedef{#1\leftbracket fig.\noexpand~\the\figno}%
\global\advance\figno by1} \fi

\def\figin{\epsfcheck\figin}\def\figins{\epsfcheck\figins}
\def\epsfcheck{\ifx\epsfbox\UnDeFiNeD
\message{(NO epsf.tex, FIGURES WILL BE IGNORED)}
\gdef\figin##1{\vskip2in}\gdef\figins##1{\hskip.5in}
\else\message{(FIGURES WILL BE INCLUDED)}%
\gdef\figin##1{##1}\gdef\figins##1{##1}\fi}
\def\DefWarn#1{}
\def\figinsert{\goodbreak\midinsert}
\def\ifig#1#2#3{\DefWarn#1\xdef#1{fig.~\the\figno}
\writedef{#1\leftbracket fig.\noexpand~\the\figno} %
\figinsert\figin{\centerline{#3}}\medskip\centerline{\vbox{\baselineskip12pt
\advance\hsize by -1truein\noindent\footnotefont{\bf
Fig.~\the\figno:} #2}}
\bigskip\endinsert\global\advance\figno by1}

\def \pa {\partial}

\def\OO{{\cal OO}}

\catcode`\@=11
\def\slash#1{\mathord{\mathpalette\c@ncel{#1}}}
\overfullrule=0pt

\def\FF{{\cal F}}
\def\GG{{\cal G}}

\def\OO{{\cal O}}

\def\underrel#1\over#2{\mathrel{\mathop{\kern\z@#1}\limits_{#2}}}

\catcode`\@=12

\def\lzbarone{{\underrel{\approx}\over{\bar{z}\rightarrow 1}}}
\def\lzbaronezone{{\underrel{\approx}\over{\bar{z}\rightarrow 1,z\rightarrow 1}}}

\def\luzero{{\underrel{\approx}\over{u \rightarrow 0}}}



\def\zbar{{\bar z}}

\def\hbar{{\bar h}}

\def\DL{{\Delta_{L}}}
\def\DH{{\Delta_{H}}}
\def\DHL{{\Delta_{HL}}}
\def\OL{{\OO_{L}}}
\def\OH{{\OO_{H}}}


\lref\ZamolodchikovGT{
  A.~B.~Zamolodchikov,
  ``Irreversibility of the Flux of the Renormalization Group in a 2D Field Theory,''
JETP Lett.\  {\bf 43}, 730 (1986), [Pisma Zh.\ Eksp.\ Teor.\ Fiz.\  {\bf 43}, 565 (1986)]..
}

\lref\KomargodskiVJ{
  Z.~Komargodski and A.~Schwimmer,
  ``On Renormalization Group Flows in Four Dimensions,''
JHEP {\bf 1112}, 099 (2011).
[arXiv:1107.3987 [hep-th]].
}

\lref\HartmanLFA{
  T.~Hartman, S.~Jain and S.~Kundu,
  ``Causality Constraints in Conformal Field Theory,''
JHEP {\bf 1605}, 099 (2016).
[arXiv:1509.00014 [hep-th]].
}

\lref\LiITL{
  D.~Li, D.~Meltzer and D.~Poland,
  ``Conformal Collider Physics from the Lightcone Bootstrap,''
JHEP {\bf 1602}, 143 (2016).
[arXiv:1511.08025 [hep-th]].
}

\lref\HartmanDXC{
  T.~Hartman, S.~Jain and S.~Kundu,
  ``A New Spin on Causality Constraints,''
JHEP {\bf 1610}, 141 (2016).
[arXiv:1601.07904 [hep-th]].
}

\lref\FaulknerMZT{
  T.~Faulkner, R.~G.~Leigh, O.~Parrikar and H.~Wang,
  ``Modular Hamiltonians for Deformed Half-Spaces and the Averaged Null Energy Condition,''
JHEP {\bf 1609}, 038 (2016).
[arXiv:1605.08072 [hep-th]].
}

\lref\HofmanAR{
  D.~M.~Hofman and J.~Maldacena,
  ``Conformal collider physics: Energy and charge correlations,''
JHEP {\bf 0805}, 012 (2008).
[arXiv:0803.1467 [hep-th]].
}

\lref\KologluMFZ{
  M.~Kologlu, P.~Kravchuk, D.~Simmons-Duffin and A.~Zhiboedov,
  ``The light-ray OPE and conformal colliders,''
[arXiv:1905.01311 [hep-th]].
}

\lref\KulaxiziDXO{
  M.~Kulaxizi, G.~S.~Ng and A.~Parnachev,
  ``Black Holes, Heavy States, Phase Shift and Anomalous Dimensions,''
SciPost Phys.\  {\bf 6}, 065 (2019).
[arXiv:1812.03120 [hep-th]].
}

\lref\CollierEXN{
  S.~Collier, Y.~Gobeil, H.~Maxfield and E.~Perlmutter,
  ``Quantum Regge Trajectories and the Virasoro Analytic Bootstrap,''
JHEP {\bf 1905}, 212 (2019).
[arXiv:1811.05710 [hep-th]].
}

\lref\BeccariaSHQ{
  M.~Beccaria, A.~Fachechi and G.~Macorini,
  ``Virasoro vacuum block at next-to-leading order in the heavy-light limit,''
JHEP {\bf 1602}, 072 (2016).
[arXiv:1511.05452 [hep-th]].
}

\lref\CornalbaXK{
  L.~Cornalba, M.~S.~Costa, J.~Penedones and R.~Schiappa,
  ``Eikonal Approximation in AdS/CFT: From Shock Waves to Four-Point Functions,''
JHEP {\bf 0708}, 019 (2007).
[hep-th/0611122].
}
\lref\CornalbaXM{
  L.~Cornalba, M.~S.~Costa, J.~Penedones and R.~Schiappa,
  ``Eikonal Approximation in AdS/CFT: Conformal Partial Waves and Finite N Four-Point Functions,''
Nucl.\ Phys.\ B {\bf 767}, 327 (2007).
[hep-th/0611123].
}
\lref\CornalbaZB{
  L.~Cornalba, M.~S.~Costa and J.~Penedones,
  ``Eikonal approximation in AdS/CFT: Resumming the gravitational loop expansion,''
JHEP {\bf 0709}, 037 (2007).
[arXiv:0707.0120 [hep-th]].
}

\lref\FitzpatrickDLT{
  A.~L.~Fitzpatrick and J.~Kaplan,
  ``Conformal Blocks Beyond the Semi-Classical Limit,''
JHEP {\bf 1605}, 075 (2016).
[arXiv:1512.03052 [hep-th]].
}
\lref\FitzpatrickDM{
  A.~L.~Fitzpatrick and J.~Kaplan,
  ``Unitarity and the Holographic S-Matrix,''
JHEP {\bf 1210}, 032 (2012).
[arXiv:1112.4845 [hep-th]].
}
\lref\FitzpatrickZHA{
  A.~L.~Fitzpatrick, J.~Kaplan and M.~T.~Walters,
  ``Virasoro Conformal Blocks and Thermality from Classical Background Fields,''
JHEP {\bf 1511}, 200 (2015).
[arXiv:1501.05315 [hep-th]].
}

\lref\FitzpatrickIVE{
  A.~L.~Fitzpatrick, J.~Kaplan, D.~Li and J.~Wang,
  ``On information loss in AdS$_{3}$/CFT$_{2}$,''
JHEP {\bf 1605}, 109 (2016).
[arXiv:1603.08925 [hep-th]].
}
\lref\FitzpatrickMJQ{
  A.~L.~Fitzpatrick and J.~Kaplan,
  ``On the Late-Time Behavior of Virasoro Blocks and a Classification of Semiclassical Saddles,''
JHEP {\bf 1704}, 072 (2017).
[arXiv:1609.07153 [hep-th]].
}

\lref\FitzpatrickYX{
  A.~L.~Fitzpatrick, J.~Kaplan, D.~Poland and D.~Simmons-Duffin,
  ``The Analytic Bootstrap and AdS Superhorizon Locality,''
JHEP {\bf 1312}, 004 (2013).
[arXiv:1212.3616 [hep-th]].
}

\lref\HofmanAWC{
  D.~M.~Hofman, D.~Li, D.~Meltzer, D.~Poland and F.~Rejon-Barrera,
  ``A Proof of the Conformal Collider Bounds,''
JHEP {\bf 1606}, 111 (2016).
[arXiv:1603.03771 [hep-th]].
}

\lref\KomargodskiEK{
  Z.~Komargodski and A.~Zhiboedov,
  ``Convexity and Liberation at Large Spin,''
JHEP {\bf 1311}, 140 (2013).
[arXiv:1212.4103 [hep-th]].
}
\lref\KomargodskiGCI{
  Z.~Komargodski, M.~Kulaxizi, A.~Parnachev and A.~Zhiboedov,
  ``Conformal Field Theories and Deep Inelastic Scattering,''
Phys.\ Rev.\ D {\bf 95}, no. 6, 065011 (2017).
[arXiv:1601.05453 [hep-th]].
}

\lref\PolandEPD{
  D.~Poland, S.~Rychkov and A.~Vichi,
  ``The Conformal Bootstrap: Theory, Numerical Techniques, and Applications,''
[arXiv:1805.04405 [hep-th]].
}

\lref\SimmonsDuffinGJK{
  D.~Simmons-Duffin,
  ``TASI Lectures on the Conformal Bootstrap,''
[arXiv:1602.07982 [hep-th]].
}

\lref\FitzpatrickVUA{
  A.~L.~Fitzpatrick, J.~Kaplan and M.~T.~Walters,
  ``Universality of Long-Distance AdS Physics from the CFT Bootstrap,''
JHEP {\bf 1408}, 145 (2014).
[arXiv:1403.6829 [hep-th]].
}

\lref\FitzpatrickZQZ{
  A.~L.~Fitzpatrick and K.~W.~Huang,
  ``Universal Lowest-Twist in CFTs from Holography,''
[arXiv:1903.05306 [hep-th]].
}

\lref\CostaCB{
  M.~S.~Costa, V.~Goncalves and J.~Penedones,
  ``Conformal Regge theory,''
JHEP {\bf 1212}, 091 (2012).
[arXiv:1209.4355 [hep-th]].
}

\lref\KulaxiziTKD{
  M.~Kulaxizi, G.~S.~Ng and A.~Parnachev,
  ``Subleading Eikonal, AdS/CFT and Double Stress Tensors,''
[arXiv:1907.00867 [hep-th]].
}

\lref\KarlssonQFI{
  R.~Karlsson, M.~Kulaxizi, A.~Parnachev and P.~Tadi\' c,
  ``Black Holes and Conformal Regge Bootstrap,''
[arXiv:1904.00060 [hep-th]].
}

\lref\FitzpatrickEFK{
  A.~L.~Fitzpatrick, K.~W.~Huang and D.~Li,
  ``Probing Universalities in $d>2$ CFTs: from Black Holes to Shockwaves,''
[arXiv:1907.10810 [hep-th]].
}

\lref\FitzpatrickHFC{
  A.~L.~Fitzpatrick, J.~Kaplan, M.~T.~Walters and J.~Wang,
  ``Hawking from Catalan,''
JHEP {\bf 1605}, 069 (2016).
[arXiv:1510.00014 [hep-th]].
}

\lref\ChenNVBIP{
  H.~Chen, C.~Hussong, J.~Kaplan and D.~Li,
  ``A Numerical Approach to Virasoro Blocks and the Information Paradox,''
[arXiv:1703.09727 [hep-th]].
}

\lref\AsplundRE{
  C.~T.~Asplund, A.~Bernamonti, F.~Galli and T.~Hartman,
  ``Holographic Entanglement Entropy from 2d CFT: Heavy States and Local Quenches,''
  JHEP {\bf 2015}, 171 (2015).
  [arXiv:1410.1392 [hep-th]].
}

\lref\ChenEE{
  B.~Chen and J.~Wu,
  ``Holographic Entanglement Entropy For a Large Class of States in 2D CFT,''
  [arXiv:1605.06753 [hep-th]].
}

\lref\ChenESC{
  B.~Chen, J.~Wu and J.~Zhang,
  ``Holographic Description of 2D Conformal Block in Semi-classical Limit,''
  [arXiv:1609.00801 [hep-th]].
}

\lref\HijanoQJA{
  E.~Hijano, P.~Kraus, E.~Perlmutter and R.~Snively,
  ``Semiclassical Virasoro blocks from AdS$_{3}$ gravity,''
JHEP {\bf 1512}, 077 (2015).
[arXiv:1508.04987 [hep-th]].
}

\lref\HartmanEE{
  T.~Hartman, 
  ``Entanglement Entropy at Large Central Charge,''
[arXiv:1303.6955 [hep-th]].
}

\lref\AnousBHC{
  T.~Anous, T.~Hartman, A.~Rovai and J.~Sonner, 
  ``Black Hole Collapse in the 1/c Expansion,''
[arXiv:1603.04856 [hep-th]].
}

\lref\LiHOPE{
  Y.~Li, Z.~Mai and H.~Lu, 
  ``Holographic OPE Coefficients from AdS Black Holes with Matters,''
[arXiv:1905.09302 [hep-th]].
}

\lref\CaputaQE{
  P.~Caputa, J.~Simon, A.~Stikonas and T.~Takayanagi, 
  ``Quantum Entanglement of Localized Excited States at Finite Temperature,''
[arXiv:1410.2287 [hep-th]].
}

\lref\RychkovIQZ{
  S.~Rychkov,
  ``EPFL Lectures on Conformal Field Theory in D$\geq$3 Dimensions,''
[arXiv:1601.05000 [hep-th]].
}

\lref\RattazziPE{
  R.~Rattazzi, V.~S.~Rychkov, E.~Tonni and A.~Vichi,
  ``Bounding scalar operator dimensions in 4D CFT,''
JHEP {\bf 0812}, 031 (2008).
[arXiv:0807.0004 [hep-th]].
}

\lref\HijanoRLA{
  E.~Hijano, P.~Kraus and R.~Snively,
  ``Worldline approach to semi-classical conformal blocks,''
JHEP {\bf 1507}, 131 (2015).
[arXiv:1501.02260 [hep-th]].
}

\lref\FaulknerHLL{
  T.~Faulkner and H.~Wang,
  ``Probing beyond ETH at large $c$,''
JHEP {\bf 1806}, 123 (2018).
[arXiv:1712.03464 [hep-th]].
}

\lref\FaulknerYIA{
  T.~Faulkner,
  ``The Entanglement Renyi Entropies of Disjoint Intervals in AdS/CFT,''
[arXiv:1303.7221 [hep-th]].
}

\lref\CornalbaQF{
  L.~Cornalba, M.~S.~Costa and J.~Penedones,
  ``Eikonal Methods in AdS/CFT: BFKL Pomeron at Weak Coupling,''
JHEP {\bf 0806}, 048 (2008).
[arXiv:0801.3002 [hep-th]].
}
\lref\CotlerZFF{
  J.~Cotler and K.~Jensen,
  ``A theory of reparameterizations for AdS$_3$ gravity,''
JHEP {\bf 1902}, 079 (2019).
[arXiv:1808.03263 [hep-th]].
}
\lref\HuangFOG{
  K.~W.~Huang,
  ``Stress-tensor commutators in conformal field theories near the lightcone,''
Phys.\ Rev.\ D {\bf 100}, no. 6, 061701 (2019).
[arXiv:1907.00599 [hep-th]].
}

\Title{
\vbox{\baselineskip8pt
}}
{\vbox{
\centerline{Leading Multi-Stress Tensors and Conformal Bootstrap}
}}

\vskip.1in
 \centerline{
Robin Karlsson, Manuela Kulaxizi, Andrei Parnachev and Petar Tadi\' c \footnote{}{karlsson, manuela, parnachev, tadicp $@$ maths.tcd.ie  }   } \vskip.1in
\centerline{\it 
School of Mathematics, Trinity College Dublin, Dublin 2, Ireland}

\vskip.7in \centerline{\bf Abstract}{
\vskip.2in 
\noindent 
Near lightcone correlators are dominated by  operators with the lowest twist.
We consider the contributions of such leading lowest twist multi-stress tensor operators to a heavy-heavy-light-light correlator in a CFT 
of any even dimensionality with a large central charge.
An infinite number of such  operators contribute, but  their sum is described by a simple ansatz.
We show that the coefficients in this ansatz can be determined recursively, thereby providing an operational procedure to compute them.
This is achieved  by bootstrapping the corresponding near lightcone correlator:  conformal data for any minimal-twist determines that for the higher
minimal-twist and so on.
To illustrate this procedure in four spacetime dimensions we determine the contributions of  double- and triple-stress tensors. 
We compute the OPE coefficients; whenever  results are available in the literature, we observe complete agreement. 
We also compute the contributions of  double-stress tensors in six spacetime dimensions and determine the corresponding OPE coefficients.
In all cases the results are consistent with the exponentiation of the near lightcone correlator.
This is similar to the situation in two spacetime dimensions for the Virasoro vacuum block.
}

\Date{September 2019}

\listtoc\writetoc
\vskip 1.57in \noindent

\eject
\newsec{Introduction and Summary}

\noindent 

\subsec{Introduction}
\noindent The two-point function of the stress tensor in Conformal Field Theories is proportional to a single parameter, the central charge $C_T$.
It generally serves as a measure of the number of degrees of freedom in the theory.
In two spacetime dimensions this statement can be made precise:\ one can define a c-function which monotonically decreases along  Renormalization Group
flows and reduces to the central charge at conformal fixed points \ZamolodchikovGT.
In four spacetime dimensions the situation is a bit more subtle and it is the $a$-coefficient in the conformal anomaly which necessarily  satisfies $a_{IR}\leq a_{UV}$ \KomargodskiVJ.
Nevertheless, in any unitary conformal field theory $a$ and $C_T$ can only differ by a number of $\OO(1)$ (see \HofmanAR\ for the original argument and 
\refs{\HartmanLFA\LiITL\KomargodskiGCI\HartmanDXC\HofmanAWC\FaulknerMZT-\KologluMFZ}
for more recent field theoretic proofs.)
Hence, to consider the limit of  infinite number of degrees of freedom one needs to take $C_T$ to infinity.

In two spacetime dimensions conformal symmetry is described by the infinite-dimensional Virasoro algebra.
This symmetry strongly constrains  correlators, especially when combined with the $C_T\to \infty$ limit.
Of particular interest is the ``heavy-heavy-light-light'' correlator, which involves two ``heavy'' operators with conformal dimension $\Delta_H \sim C_T$
and two ``light'' operators with conformal dimension $\DL \sim \OO(1)$. In this case the contribution of the identity operator and all its Virasoro descendants is known as the 
Virasoro vacuum block and has been calculated in several  ways \refs{\FitzpatrickVUA\FitzpatrickZHA\HijanoRLA\HijanoQJA\FitzpatrickHFC\CotlerZFF-\CollierEXN}. 
The Virasoro vacuum block (and finite $C_T$ corrections to it) is instrumental in a variety of settings,
such as e.g. the problem of information loss
 \refs{\FitzpatrickDLT\FitzpatrickIVE\AnousBHC\FitzpatrickMJQ\ChenNVBIP-\FaulknerHLL} and properties of the Renyi and entanglement entropies \refs{\AsplundRE\CaputaQE\ChenEE-\ChenESC} 
 (see also  \refs{\HartmanEE,\FaulknerYIA} for the original applications of 
 large $C_T$ correlators in this context).

The  heavy-heavy-light-light Virasoro vacuum block  exponentiates
\eqn\virvac{ \langle \OO_H(\infty) \OL(1) \OL(z) \OO_H(0) \rangle \sim e^{\DL \FF(\mu;z)}, }
with $\FF$  a known function which admits an expansion in powers of $\mu \sim \Delta_H/C_T$
\eqn\fexp{   \FF(\mu;z) =\sum_k \mu^k\FF^{(k)}(z)      .}
One can  consider contributions of various quasi-primaries made out of the stress tensor to $\FF^{(k)}$.
At $k=1$ the only such quasi-primary is the stress tensor itself, while for $k=2$ one needs to sum an infinite number
of quasi-primaries quadratic in the stress tensor (double-stress operators) and labeled by spin.
The situation is similar for all other values of $k$.
It is possible to compute the OPE coefficients of the corresponding quasi-primaries, starting from the known result for the Virasoro vacuum block. Interestingly, at each order in $\mu$,  $\FF^{(k)}$ can be written  as a sum of particular terms \KulaxiziDXO\foot{Similar expressions in a slightly different context appeared in \BeccariaSHQ.}
\eqn\fktwod{\FF^{(k)}(z) = \sum_{\{i_p\}}b_{i_1 ... i_k} f_{i_1}(z) ...  f_{i_k}(z), \qquad \sum_{p=1}^k i_p = 2k,}
where $f_a(z)  = (1-z)^a  {}_2 F_1(a,a,2 a,1-z)$. 

It is an interesting question whether a similar structure appears when the number of spacetime dimensions $d$ is greater than two.
Unlike in two spacetime dimensions, in addition to  spin, multi-stress tensor operators are also labeled by their twist.
An interesting subset of multi-stress tensor operators is comprised out of those with minimal twist.
These operators dominate in the lightcone limit over those of higher twist.
In \KulaxiziTKD\ an expression for the OPE coefficients of two scalars and minimal-twist double-stress tensor operators in $d=4$ was obtained, 
and the sum was performed to obtain a remarkably simple expression for the near lightcone $\OO(\mu^2)$ term in the heavy-heavy-light-light correlator. It was shown to have a similar form to \fktwod. 
One may now wonder if the minimal-twist multi-stress tensor part of the correlator in higher dimensions exponentiates 
\eqn\hd{ \langle \OO_H(\infty) \OL(1) \OL(z,\zbar) \OO_H(0) \rangle\big |_{\rm multi-stress\; tensors} \sim e^{\DL \FF(\mu;z,\zbar)}, }
and whether $\FF(\mu;z,\zbar)$ can be expressed as
\eqn\fexp{   \FF(\mu;z,\zbar) =\sum_k \mu^k\FF^{(k)}(z,\zbar),}
with
\eqn\fkhd{\FF^{(k)}(z,\zbar) = (1-\zbar)^{k({{d-2}\over 2})}\sum_{\{i_p\}}b_{i_1 ... i_k} f_{i_1}(z) ...  f_{i_k}(z), \qquad \sum_{p=1}^k i_p = k\left({{d+2}\over 2}\right),}
and $d$ an even number. 

In this paper we investigate this further. We start by assuming that the multi-stress tensor sector of the heavy-heavy-light-light correlator
in the near lightcone regime $\bar z \to 1$ admits an expansion in $\mu$ 
\eqn\ghd{ \langle \OO_H(\infty) \OL(1) \OL(z,\zbar) \OH(0) \rangle\big |_{\rm multi-stress\; tensors}    \sim  \sum_k \mu^k\GG^{(k)}(z,\zbar)   ,}
where each coefficient function $\GG^{(k)}(z,\zbar)$ takes a particular form:
\eqn\gkhd{\GG^{(k)}(z,\zbar) = {(1-\zbar)^{k({{d-2}\over 2})}\over [(1-z)(1-\zbar)]^{\DL}}\sum_{\{i_p\}}a_{i_1 ... i_k} f_{i_1}(z) ...  f_{i_k}(z), \qquad \sum_{p=1}^k i_p = k\left({{d+2}\over 2}\right).}
We subsequently use this ansatz to compute the contributions of the multi-stress tensor operators to the near lightcone correlator
and extract the corresponding OPE coefficients. 

For even $d$, the hypergeometric functions in \gkhd\ reduce to terms which contain at most one power of $\log(z)$ each. 
Their products contain multi-logs whose coefficients turn out to be rational functions of $z$. 
We use the conformal bootstrap approach initiated in \RattazziPE\ (for a review and references see eg. \refs{\RychkovIQZ\SimmonsDuffinGJK-\PolandEPD}) to relate these functions to the
anomalous dimensions and OPE coefficients of the heavy-light double-twist operators in the cross channel.
The ansatz \gkhd\ has just a few coefficients at any finite $k$ which can be determined completely
from the cross-channel data derived using the $(k-1)$th term.
This is related to the fact that all the $\log^m (z)$ terms with $2\leq m \leq k$ are completely determined by the anomalous
dimensions and OPE coefficients at $\OO(\mu^{k-1})$.
At each step, we obtain an overconstrained system of equations solved by the same set of $a_{i_1... i_k}$. 
This provides strong support to the ansatz \fkhd. We then proceed to derive the OPE coefficients of the multi-stress tensor operators with two light scalars from our result.
In practice, we complete this program to $\OO(\mu^3)$ in $d=4$ and to $\OO(\mu^2)$ in $d=6$. However the procedure outlined can be easily generalised to arbitrary order in $\mu$ and any even  $d$.

In \FitzpatrickZQZ\ the authors considered holographic CFTs dual to gravitational theories defined by the
 Einstein-Hilbert Lagrangian plus higher derivative terms and a scalar field minimally coupled to gravity in $AdS_{d+1}$.
Interpreting the scalar propagator in an asymptotically $AdS_{d+1}$ black hole background as a heavy-heavy-light-light four point function, enabled the authors of \FitzpatrickZQZ\ to 
extract the OPE coefficients of a few multi-stress tensor operators from holography (see also \refs{\LiHOPE\FitzpatrickEFK-\HuangFOG} for related work). 
Ref. \FitzpatrickZQZ\ also argued that the OPE coefficients of the leading, minimal-twist multi-stress operators are universal -- they do not
depend on the higher derivative terms in the Lagrangian. Their results  agree
with the general expressions obtained in this paper, upon substitution of the relevant quantum numbers.
We do not use holography in our work;
our major assumption is \gkhd. It would be interesting to see what is the regime of applicability of our results.


\subsec{Summary of the results}

\noindent 
In this paper we argue that for a large class of CFTs (including holographic CFTs)  in 
even $d$, the contribution of minimal-twist multi-stress tensors to the correlator in the lightcone limit 
can be written as a sum of products of certain hypergeometric functions. To be explicit, let us define functions $f_{a}(z)$ as
\eqn\defone{f_{a}(z)=(1-z)^{a}{}_{2}F_{1}(a,a,2a;1-z).
}
\noindent The stress tensor contribution to the correlator in the lightcone limit is given in any dimension $d$ by 
\eqn\ordermu{
    \GG^{(1)}(z,\zbar) \lzbarone {(1-\zbar)^{d-2\over 2}\over[(1-z)(1-\zbar)]^{\DL}}{\DL\Gamma({d\over 2}+1)^2\over 4\Gamma(d+2)}f_{d+2\over 2}(z).
}
At $\OO(\mu^2)$ the contribution from twist-four double-stress tensor operators in $d=4$ is 
\eqn\secondCorrIntro{\eqalign{
  \GG^{(2)}&(z,\zbar)\lzbarone {(1-\zbar)^2\over [(1-z)(1-\zbar)]^{\DL}}\left({\DL\over 28800(\DL-2)}\right)\times\cr
      \Big(&(\DL-4)(\DL-3)f_3^2(z)+{15\over 7}(\DL-8)f_2(z)f_4(z)+{40\over 7}(\DL+1)f_1(z)f_5(z)\Big) \,.
}}  
This result  agrees with the expression obtained by different methods in \KulaxiziTKD. 

The contribution from twist-six triple-stress tensors in the lightcone limit in $d=4$ at order $\OO(\mu^{3})$ is 
\eqn\anzThirdone{\eqalign{
  &\GG^{(3)}(z,\zbar)\lzbarone {{(1-\zbar)^{3}}\over{[(1-z)(1-\zbar)]^{\Delta_{L}}}} \Big(a_{117} f_{1}(z)^2 f_7(z) +a_{126}f_1(z)f_2(z)f_6(z) \cr 
  &  +a_{135}f_1(z)f_3(z)f_5(z) +a_{225}f_2(z)^2f_5(z)+a_{234}f_2(z)f_3(z)f_4(z)+a_{333}f_3(z)^3\Big),
}}
\noindent where coefficients $a_{ijk}$ are given by (3.18).

Furthermore, from \anzThirdone\ and (3.18), we find the OPE coefficients of twist-six triple-stress tensor operators as a finite sum (for details see Section 3.4). 
Two such OPE coefficients for twist-6 triple-stress tensors were calculated holographically in \FitzpatrickZQZ\ and agree with our results.

The contribution from twist-eight double-stress tensors to the correlator in the lightcone limit in $d=6$ at order $\OO(\mu^{2})$ is 
\eqn\anzSixDone{\eqalign{
  \GG^{(2)}(z,\zbar)&\lzbarone {{(1-\zbar)^{4}}\over{[(1-z)(1-\zbar)]^{\Delta_{L}}}}\times\cr
  &\Big(a_{13}f_1(z)f_7(z) +a_{26}f_2(z)f_6(z) +  a_{35}f_3(z)f_5(z)+a_{44}f_4(z)^2\Big),
}}
\noindent where $a_{mn}$ are given by (4.7).
\noindent Using \anzSixDone\ and (4.7) we find the OPE coefficients for operators of type $:T_{\mu \nu}\partial_{\lambda_{1}}\ldots\partial_{\lambda_{2l}}T_{\alpha\beta}:$ in $d=6$ to be equal to:

\eqn\opeCoeffSixDTTo{
   P^{(HH,LL)}_{8,s} = \mu^2{c\DL\over (\DL-3)(\DL-4)}(a_3\Delta_L^3+a_2\Delta_L^2+a_1\DL+a_0),
}
where $c$ and $a_{m}$, given  by (4.15),  are functions of the total spin  $s=4+2l$ .

 In general we propose that the contribution from minimal-twist multi-stress tensor operators to the correlator in even $d$ at $\OO(\mu^k)$ in the lightcone limit takes the form
\eqn\genan{\GG^{(k)}(z,\zbar)\lzbarone{{(1-\zbar)^{k({d\over 2}-1)}}\over{[(1-z)(1-\zbar)]^{\Delta_{L}}}} \sum_{\{i_p\}}a_{i_1 ... i_k} f_{i_1}(z) ...  f_{i_k}(z), \qquad \sum_{p=1}^k i_p = k\Big({d+2\over 2}\Big),
}
\noindent where the sum goes over all sets of $\{i_{p}\}$ with $i_p\leq i_{p+1}$ and $a_{i_{1} ... i_{k}}$ coefficients that need to be fixed.\foot{One only needs to sum  the linearly independent products of functions $f_{a}$.}

We also check that the stress tensor sector of the near lightcone correlator exponentiates
\eqn\expon{\langle \OH(x_4)\OL(1)\OL(z,\zbar)\OH(0)\rangle |_{\rm multi-stress\;tensors}\lzbarone {{1}\over{[(1-z)(1-\zbar)]^{\Delta_{L}}}}e^{\DL\FF(\mu;z,\zbar)}
,}
\noindent where $\FF(\mu;z,\zbar)$ is a rational function of $\DL$ that remains $\OO(1)$ as $\DL\to\infty$. We explicitly verify this up to $\OO(\mu^3)$ in $d=4$ and $\OO(\mu^2)$ in $d=6$.

\subsec{Outline}

\noindent The rest of the paper is organized as follows. In Section 2, we establish notation and write general expressions for the heavy-heavy-light-light correlator in both the direct channel (T-channel) and the cross channel (S-channel). We further write down the stress tensor contribution to the correlator in the lightcone limit in arbitrary spacetime dimensions $d$. In Section 3, we find the contribution of minimal-twist double- and triple-stress tensor operators in $d=4$ in the lightcone limit. We show that this contribution exponentiates and we write an expression for the OPE coefficients of minimal-twist triple-stress tensors of spin $s$ with scalar operators, in the form of a finite sum. 
In Section 4, we repeat this program up to $\OO(\mu^2)$ in $d=6$. Again we confirm exponentiation and we find a closed form expression for the OPE coefficients of minimal-twist double-stress tensors of arbitrary spin with scalar operators. We discuss our results in Section 5.

\newsec{Review of heavy-heavy-light-light correlator in the lightcone limit}
\noindent Below we review the setup of a heavy-heavy-light-light correlator with focus on its behaviour in the lightcone limit. 
We mostly follow \refs{\KulaxiziDXO,\KarlssonQFI,\KulaxiziTKD}. 

The object that we study is a four-point function of pairwise identical scalars 
	$\langle \OH(x_4)\OL(x_3)\OL(x_2)\OH(x_1)\rangle$.
Here $\OH$ and $\OL$ are scalar operators with scaling dimension $\DH\propto \OO(C_T)$ and $\DL\propto \OO(1)$, with $C_T\gg 1$ the central charge.

Using conformal transformations we define the stress tensor sector of the correlator by
\eqn\Corrdef{
	\GG(z,\zbar) = \lim_{x_4\to\infty} x_4^{2\DH}\langle \OH(x_4)\OL(1)\OL(z,\zbar)\OH(0)\rangle \Big|_{\rm multi-stress\; tensors} ,
} 
where $z$ and $\zbar$ are the usual cross-ratios\foot{Note  that $(u,v)$ are exchanged compared to the more common convention.}
\eqn\cross{\eqalign{
	u &= (1-z)(1-\zbar) = {x_{14}^2 x_{23}^2\over x_{13}^2x_{24}^2},\cr
	v &= z\zbar = {x_{12}^2 x_{34}^2 \over x_{13}^2 x_{24}^2}.
}}
In \Corrdef\ the``multi-stress tensor" subscript stands to indicate the contribution of the identity and
all multi-stress tensor operators.

The  correlator $\GG(z,\zbar)$ can be expanded in the ``T-channel'' $\OL(1)\times\OL(z,\zbar)\to \OO_{\tau,s}$ as\foot{For reasons of convenience, here and in the rest of the paper we refer to $\GG(z,\zbar)$ as the correlator; the reader  should keep in mind that $\GG(z,\zbar)$ is not the full correlator but only its stress tensor sector, as defined in \Corrdef.} 
\eqn\TChannel{
	\GG(z,\zbar) = [(1-z)(1-\zbar)]^{-\DL}\sum_{\OO_{\tau,s}} P^{(HH,LL)}_{\OO_{\tau,s}} g_{\tau,s}^{(0,0)}(1-z,1-\zbar),
}
where $\tau=\Delta-s$ and $s$ denote the twist and spin of the exchanged operator, respectively, and $g_{\tau,s}^{(0,0)}(z,\zbar)$ the conformal block of the primary operator $\OO_{\tau,s}$. 
Moreover, $P^{(HH,LL)}_{\OO_{\tau,s}}$ are defined as
\eqn\OPESquar{
	P^{(HH,LL)}_{\OO_{\tau,s}} = \left(-{1\over 2}\right)^s\lambda_{\OH\OH\OO_{\tau,s}}\lambda_{\OL\OL\OO_{\tau,s}},
}
where $\lambda_{\OL\OL\OO}$ and $\lambda_{\OH\OH\OO}$ denote the respective OPE coefficients. 

We will mainly be interested in the lightcone limit defined by $u\ll 1$ or equivalently $\zbar\to 1$. In this limit the T-channel expansion \TChannel\ is dominated by minimal-twist operators as follows from the behaviour of the conformal blocks
\eqn\LCBehaviour{
	\GG(u, v) \luzero  u^{-\DL}\sum_{\OO_{\tau,s}}P^{(HH,LL)}_{\OO_{\tau,s}} u^{\tau\over 2}(1-v)^{-{\tau \over 2}}f_{{\tau\over 2}+ s}(v),
}
where $\tau=\Delta-s$ is the twist and
\eqn\defFfunc{
	f_{{\tau\over 2}+ s}(v) = (1-v)^{{\tau\over 2}+ s}{}_2F_1\Big({{\tau\over 2}+ s},{{\tau\over 2}+ s},\tau+2s,1-v\Big)
}
is a $SL(2;R)$ conformal block.

For any CFT in $d>2$ the leading contribution in the lightcone limit comes from the exchange of the identity operator with twist $\tau=0$. Another operator present in any unitary CFT is the stress tensor with twist $\tau=d-2$. 
Its contribution to the correlator is completely fixed by a Ward identity  and 
\eqn\OPEcoeffStress{
	P^{(HH,LL)}_{T_{\mu\nu}} = \mu{\DL\over 4}{\Gamma({d\over 2}+1)^2\over \Gamma(d+2)},
}
where 
\eqn\mudef{
	\mu := {4\Gamma(d+2)\over(d-1)^2 \Gamma({d\over 2})^2}{\DH\over C_T}.
}
As explained in \refs{\KulaxiziDXO}, the correlator admits a natural perturbative expansion in $\mu$,
\eqn\Gkdef{\GG(z,\zbar)=\sum_k \mu^k\GG^{(k)}(z,\zbar)\,.}

Using \LCBehaviour\ and \OPEcoeffStress, we find the following contribution due to the stress tensor at $\OO(\mu)$
\eqn\StressTensorCont{
  	\GG^{(1)}(z,\zbar)\lzbarone {(1-\zbar)^{d-2\over 2}\over[(1-z)(1-\zbar)]^{\DL}}{\DL\Gamma({d\over 2}+1)^2\over 4\Gamma(d+2)}(1-z)^{d+2\over 2}{}_2F_1\Big({d+2\over 2},{d+2\over 2};d+2;1-z\Big).
}

Let us study the correlator perturbatively in $\mu$ in the lightcone limit. At $k$-th order in that expansion we expect contributions from minimal-twist multi-stress tensor operators of the schematic form
 $[T^k]_{\tau,s}=:T_{\mu_1\nu_1}\ldots\pa_{\lambda_1}\ldots \pa_{\lambda_l}T_{\mu_k\nu_k}:$, where the minimal-twist $\tau$ and spin $s$ of these operators are given by 
\eqn\twistandspin{\eqalign{
  \tau &= k(d-2),\cr 
  s &= 2k+l
}}  
and $l$ an even integer denoting the number of uncontracted derivatives.
We moreover define the product of OPE coefficients for minimal-twist operators at order $k$ as 
\eqn\OPEDefMultiStress{
	P^{(HH,LL)}_{[T^k]_{\tau,s}} = \mu^k P^{(HH,LL);(k)}_{\tau,s}.
}
Compared to the $k=1$ case, there exists an infinite number of minimal-twist multi-stress tensor operators for each value of $k>1$. 
To obtain their contribution to the correlator in the lightcone limit, we thus have to sum over all these operators.  

The correlator can likewise be expanded in the ``S-channel'' $\OL(z,\zbar)\times \OH(0)\to\OO_{\tau',s'}$ as 
\eqn\sCh{
	\GG(z,\zbar) = (z\zbar)^{-{1\over 2}(\DH+\DL)}\sum_{\OO_{\tau',s'}}P^{(HL,HL)}_{\OO_{\tau',s'}}g_{\tau',s'}^{(\DHL, -\DHL)}(z,\zbar).
}
where $P^{(HL,HL)}_{\OO_{\tau',s'}}$ are the products of the  corresponding OPE coefficients and $\DHL=\Delta_H-\Delta_L$.
Operators contributing in the S-channel are ``heavy-light double-twist operators'' \refs{\KulaxiziDXO,\KarlssonQFI}\foot{This the analogue of light-light double-twist operators that are present in the cross channel of $\langle \OO_{1}\OO_2\OO_2\OO_1\rangle$, with $\OO_1$ and $\OO_2$ both light, in any CFT \refs{\FitzpatrickYX,\KomargodskiEK}.} that can be schematically written as
 $[\OH\OL]_{n,l}=:\OH\pa^{2n}\pa^{l}\OL:$, with scaling dimension $\Delta_{n,l}=\DH+\DL+2n+l+\gamma(n,l)$ and spin $l$. 
 In the $\DH\to\infty$ limit the  $d=4$ blocks  are given by
\eqn\SblocksFourD{
  g^{(\DHL, -\DHL)}_{\DH+\DL+2n+\gamma,l}(z,\bar z) \approx {(z\zbar)^{{1\over 2}(\DH+\DL+2n+\gamma)}\over \zbar-z}\left(\zbar^{l+1}-z^{l+1}\right),
}
and similarly  in $d=6$
\eqn\SblockssSixD{
  g_{\DH+\DL+2n+\gamma,l}^{(\DHL, -\DHL)}(z,\zbar)\approx{(z\zbar)^{{1\over 2}(\DH+\DL+2n+\gamma(n,l))}\over (\zbar-z)^3}\left(\zbar^{l+3}-{l+3\over l+1}\zbar^{l+2}z^{1}-(z\leftrightarrow \zbar)\right).
}

The anomalous dimensions $\gamma(n,l)$ admit an expansion in $\mu$ 
\eqn\anomdimexp{
  \gamma(n,l)= \sum_{k=1}^\infty \mu^k \gamma^{(k)}_{n,l}.
}
Likewise,  we expand the product of the OPE coefficients of the double-twist operators as 
\eqn\OPECoeffSCH{
	P^{(HL,HL)}_{n,l} = P^{(HL,HL); {\rm MFT}}_{n,l}\sum_{k=0}^{\infty}\mu^k P^{(HL,HL); (k)}_{n,l},
}
with $P^{(HL,HL); (0)}_{n,l}=1$. The zeroth order OPE coefficients $ P^{(HL,HL); {\rm MFT}}_{n,l}$ in the S-channel are those of Mean Field Theory found in \FitzpatrickDM. In the limit $\DH\to\infty$ they are given by 
\eqn\OPEHeavy{
	 P^{(HL,HL); {\rm MFT}}_{n,l} \approx {(\DL-d/2+1)_n (\DL)_{l+n}\over n!\,l!\,(l+d/2)_n },
}
where $(a)_n$ denotes the Pochhammer symbol. For large $l$ we further approximate \OPEHeavy\ by 
\eqn\OPEHeavyLargel{
	 P^{(HL,HL); {\rm MFT}}_{n,l}\approx {l^{\DL-1} (\DL-{d\over 2}+1)_n\over n!\,\Gamma(\DL)}.
}
To reproduce the correct singularities manifest in the T-channel one has to sum over infinitely many heavy-light double-twist operators with $l\gg 1$. 
For such operators the dependence 
of the OPE data on the spin $l$ for $l\gg 1$ is\foot{This behavior in the large $l$ limit is different from that of the OPE data
of light-light double-twist operators \refs{\FitzpatrickYX,\KomargodskiEK}.}:
\eqn\Spindep{\eqalign{
  P^{(HL,HL);(k)}_{n,l}&= {P^{(k)}_n\over l^{k(d-2)\over 2}},\cr 
  \gamma^{(k)}_{n,l}&= {\gamma^{(k)}_n\over l^{k(d-2)\over 2}}.
}}
Note that generally the OPE data in the S-channel
receives corrections needed to reproduce double-twist operators in the T-channel; however, since we are interested in the stress tensor sector we consider only contributions of the form given in \Spindep.

\newsec{Multi-stress tensors in four dimensions}
\noindent In this section we describe how to use crossing symmetry to fix the contribution of minimal-twist multi-stress tensors to the heavy-heavy-light-light correlator in $d=4$ to $\OO(\mu^3)$. The methods described generalize to other even spacetime dimensions, with the six-dimensional case to $\OO(\mu^2)$ described in Section 4.
 In principle the same technology can also be used to determine the correlator at higher orders. Moreover, the resulting expression can be decomposed into multi-stress tensor blocks of minimal-twist, allowing us at each order to read off the OPE coefficients of minimal-twist multi-stress tensors. 

The idea is to study the S-channel expansion in \sCh\ in the limit $1-\zbar\ll z \ll 1$. In this limit operators with $l\gg 1$ and low values of $n$ dominate. Expanding the conformal blocks in \SblocksFourD\ for small $\gamma(n,l)$ and $\zbar\to 1$, the blocks in $d=4$ reduce to 
\eqn\SchBlockAppx{
  (z\zbar)^{-{1\over 2}(\DH+\DL)}g_{\DH+\DL+2n+\gamma,l}^{(\DHL, -\DHL)}(z,\zbar)\,\, \lzbarone \,\,\zbar^l p(\log z ,\gamma(n,l)){z^n\over 1-z},
}
where $p(\log z, \gamma(n,l))$ is given by 
\eqn\pfunc{
  p(\log z,\gamma(n,l)) = z^{{1\over 2}\gamma(n,l)} = \sum_{j=0}^\infty {1\over j!}\left({\gamma(n,l)\log z\over 2}\right)^j. 
}
Inserting \SchBlockAppx\ into \sCh\ and converting the sum into an integral, we have the following expression for the correlator in the limit $\zbar\to 1$
\eqn\corrlcS{
  \GG(z,\zbar) \,\, \lzbarone \,\, \sum_{n=0}^\infty {z^n\over 1-z}\int_0^\infty dl P^{(HL,HL)}_{n,l}\zbar^{l}p(\log z,\gamma(n,l)). 
}
In the following we consider an expansion of \corrlcS\ around $z=0$.
The key point is to note that by expanding the anomalous dimensions and OPE coefficients, as in \anomdimexp\ and \OPECoeffSCH\ respectively, terms proportional to $z^p\log^i z$ with $i=2,3,\ldots,k$ and any $p$ at $\OO(\mu^k)$, in \corrlcS\ are completely determined in terms of OPE data at $\OO(\mu^{k-1})$. Moreover, using \Spindep\ one sees that the integral over the spin $l$ yields
\eqn\lintegral{
  \int_{0}^\infty dl l^{\DL-1-k} \zbar^{l} = {\Gamma(\DL-k)\over (-\log \bar z)^{\DL-k}}\,\,\lzbarone\,\, {\Gamma(\DL-k)\over (1-\zbar)^{\DL-k}},
}
at $\OO(\mu^k)$ in the limit $\zbar\to 1$. This correctly reproduces the expected $\zbar$ behaviour of minimal-twist multi-stress tensors in the T-channel, thus verifying  \Spindep.

We now make the following ansatz for the correlator
\eqn\AnzFourD{\GG^{(k)}(z,\zbar)\,\lzbarone \,{{(1-\zbar)^{k}}\over{[(1-z)(1-\zbar)]^{\Delta_{L}}}} \sum_{\{i_{p}\} } a_{i_{1}\ldots i_{k}} f_{i_{1}}(z)\ldots f_{i_{k}}(z),
}
where the sum goes over all sets of $\{i_{p}\}$ with $i_p$ integers and $i_p\leq i_{p+1}$ such that $\sum_{p=1}^{k} i_{p} = 3k$ and $a_{i_{1} ... i_{k}}$ coefficients that need to be fixed.
Generally $f_{a}(z)$ are given by 
\eqn\ffuncA{
  f_{a}(z)= q_{1,a}(z)+q_{2,a}(z)\log z,
} 
where $q_{(1,2),a}(z)$ are rational functions and the ansatz \AnzFourD\ at $\OO(\mu^k)$ is therefore a polynomial in $\log z$ of degree $k$. By crossing symmetry terms with $\log^{a} z$, with $2\leq a\leq k$, are determined by OPE data at $\OO(\mu^{k-1})$. This is what we will use to determine the coefficients $a_{i_{1} ... i_{p}}$.

\subsec{Stress tensor}

\noindent We start by determining the OPE data at $\OO(\mu)$. This is easily obtained by matching \corrlcS\ at $\OO(\mu)$ with the stress tensor contribution \StressTensorCont. Explicitly, multiplying both channels by $(1-z)$ we have at $\OO(\mu)$
\eqn\Expandrelation{
  {\DL f_3(z)\over 120[(1-z)(1-\zbar)]^{\DL-1}} = {1\over (1-\zbar)^{\DL-1}}\sum_{n=0}^\infty {\Gamma(\DL+n-1)z^n\over \Gamma(\DL)n!}\left(P^{(1)}_n+{\gamma^{(1)}_n\over 2}\log z\right).
} 
Expanding the LHS in \Expandrelation\ for $z\ll1$ we find 
\eqn\stressExpansion{\eqalign{
   {\DL/120\over [(1-z)(1-\zbar)]^{\DL-1}}f_{3}(z)&= {1\over (1-\zbar)^{\DL-1}}\Big(-{\DL\over 4}(3+\log z)\cr
   &-z{\DL\over 4}(3(\DL+1)+(\DL+5)\log z)\cr
   &-z^2{\DL\over 8}\left(3\DL(\DL+3)+(12+\DL(\DL+11))\right)\cr
   &+\OO(z^3,z^3\log z)\Big),
}}
while the RHS is given by 
\eqn\rhsMatchStresstensor{\eqalign{
  {\sum_{n=0}^\infty {\Gamma(\DL+n-1)z^n\over \Gamma(\DL)n!}(P^{(1)}_n+{\gamma^{(1)}_n\over 2}\log z)\over (1-\zbar)^{\DL-1}} &= {1\over (1-\zbar)^{\DL-1}}\Big({P^{(1)}_0+{\gamma^{(1)}_0\over 2}\log z\over \DL-1}\cr
  &+z(P^{(1)}_1+{\gamma^{(1)}_1\over 2}\log z)\cr
  &+z^2{\DL\over 2}(P^{(1)}_2+{\gamma^{(1)}_2\over 2}\log z)\cr
  &+\OO(z^3,z^3\log z)\Big).
}}
Comparing \stressExpansion\ and \rhsMatchStresstensor\ order-by-order in $z$ one finds the following OPE data
\eqn\anomdimfirstorder{\eqalign{
  \gamma^{(1)}_0 &= -{\DL(\DL-1)\over 2},\cr
  \gamma^{(1)}_1 &= -{\DL(\DL+5)\over 2},\cr
  \gamma^{(1)}_2 &= -{12+\DL(\DL+11)\over 2},
}}
which agrees with eq.\ (6.10) in \KulaxiziDXO, and the OPE coefficients
\eqn\OPEFirstOrder{\eqalign{
  P^{(1)}_0 &= -{3\DL(\DL-1)\over 4},\cr 
  P^{(1)}_1 &= -{3\DL(\DL+1)\over 4},\cr 
  P^{(1)}_2 &= -{3\DL(\DL+3)\over 4}. 
}}
It is straightforward to continue and compute the $\OO(\mu)$ OPE data in the S-channel for any value of $n$.

\subsec{Twist-four double-stress tensors}

\noindent From \AnzFourD\ we infer the following expression for the contribution due to twist-four double-stress tensors to the heavy-heavy-light-light correlator in the limit $\zbar\to 1$:
\eqn\GTwoFourD{
  \GG^{(2)}(z,\zbar)\lzbarone {(1-\zbar)^2\over [(1-z)(1-\zbar)]^{\DL}}\Big(a_{15}f_1(z)f_5(z)+a_{24}f_2(z)f_4(z)+a_{33}f_3^2(z)\Big). 
} 
By expanding \GTwoFourD\ further in the limit $z\ll 1$ and collecting terms that goes as $z^p\log^2 z$, we will match with known contributions obtained from \corrlcS.

Inserting \anomdimfirstorder\ and \OPEFirstOrder\ in the S-channel \corrlcS\ fixes terms proportional to $z^p\log^2 z$ up to $\OO(z^2\log^2 z)$. Expanding the ansatz \GTwoFourD\ and matching with the S-channel reproduces the result obtained in \KulaxiziTKD:
\eqn\secondCorr{\eqalign{
  \GG^{(2)}&(z,\zbar)\lzbarone {(1-\zbar)^2\over [(1-z)(1-\zbar)]^{\DL}}\left({\DL\over 28800(\DL-2)}\right)\times\cr
      &\Big\{(\DL-4)(\DL-3)f_3^2(z)+{15\over 7}(\DL-8)f_2(z)f_4(z)+{40\over 7}(\DL+1)f_1(z)f_5(z)\Big\}.
}} 
Using the $\OO(\mu)$ OPE data in the S-channel for $n>2$ in \stressExpansion\ and \rhsMatchStresstensor\ one gets an overconstrained system which is 
still solved by \secondCorr.
This is a strong argument in favor of the validity of our ansatz \AnzFourD.

We can now use \secondCorr\ to derive the $\OO(\mu^2)$ OPE data in the S-channel by matching terms proportional to $z^p\log^i z$ as $z\to 0$, with $i=0,1$, by comparing with \corrlcS. 
This is done  in the same way it was done for $\OO(\mu)$ OPE data in the S-channel. For example, one finds the following data for $n=0,1,2,3$:
\eqn\anomdimsecondorder{\eqalign{
  \gamma^{(2)}_0 &= -{(\DL-1)\DL(4\DL+1)\over 8},\cr
  \gamma^{(2)}_1 &= -{\DL(\DL+1)(4\DL+35)\over 8},\cr
  \gamma^{(2)}_2 &= -{(3+\DL)(68+\DL(69+4\DL))\over 8},\cr
  \gamma^{(2)}_3 &= -{(5+\DL)(204+\DL(4\DL+103))\over 8},
}} 
which agrees with Eq.\ (6.39) in \KulaxiziDXO, and for the OPE coefficients
\eqn\OPESecondOrder{\eqalign{
  P^{(2)}_0 &= {(\DL-1)\DL(-28+\DL(-145+27\DL))\over 96},\cr 
  P^{(2)}_1 &= {\DL(-596+\DL(-399+\DL(-64+27\DL)))\over 96},\cr 
  P^{(2)}_2 &= {-1248+\DL(-2252+\DL(-699+\DL(44+27\DL)))\over 96},\cr 
  P^{(2)}_3 &= {-3744+\DL(-4940+\DL(-783+\DL(152+27\DL)))\over 96}. 
}}
It is again straightforward to extract the OPE data for any value of $n$. 

\subsec{Twist-six triple-stress tensors}
\noindent We now consider the multi-stress tensor sector of the correlator at $\OO(\mu^3)$ and proceed 
similarly to the previous Section. From \AnzFourD\ we infer the following expression for the contribution due to twist-six triple-stress tensors:
\eqn\AnzThird{\eqalign{
  \GG^{(3)}(z,\zbar)\lzbarone\,\, {(1-\zbar)^3\over [(1-z)(1-\zbar)]^{\DL}}\Big(&a_{117}f_{1}^2f_7+a_{126}f_1f_2f_6+a_{135}f_1f_3f_5\cr
  +&a_{225}f_2^2f_5+a_{234}f_2f_3f_4+a_{333}f_3^3\Big),
}}
where $f_{i}=f_i(z)$ is given by \defFfunc.\foot{Note that we omitted a potential term of the form $f_1f_4^2$. This can be written in terms of $f_3^3$, $f_{1}f_3f_5$, $f_{2}^2f_5$ and $f_2f_3f_4$, as follows from: 
\eqn\equality{
  f^3_3(z) = {20\over 21}f_1(z)f_3(z)f_5(z)-{27\over 28}f_1(z)f^2_4(z)-{20\over 21}f^2_2(z)f_5(z)+{55\over 28}f_2(z)f_3(z)f_4(z).}
  } 
Taking the limit $1-\zbar\ll z \ll 1$ of \AnzThird, we fix the coefficients by matching with terms proportional to $z^p\log^2 z$ and $z^p\log^3 z$, with $p=0,1,2$ from \corrlcS. 
This requires using the OPE data of the heavy-light double-twist operators $[\OH\OL]_{n,l}$ for $n=0,1,2$ and $l\gg 1$ to $\OO(\mu^2)$, 
 given in \anomdimfirstorder, \OPEFirstOrder, \anomdimsecondorder\ and \OPESecondOrder. 

We find the following solution:
\eqn\ResultTripleFourD{\eqalign{
    a_{117} &= {5\DL(\DL+1)(\DL+2)\over 768768(\DL-2)(\DL-3)},\cr
    a_{126} &= {5\DL( 5\Delta_L^2-57\DL-50)\over 6386688(\DL-2)(\DL-3)},\cr
    a_{135} &= {\DL (2\Delta_L^2-11\DL-9)\over 1209600 (\DL-3)},\cr 
    a_{225} &= -{\DL(7\Delta_L^2-51\DL-70)\over 2903040(\DL-2)(\DL-3)},\cr 
    a_{234} &= {\DL(\DL-4)(3\Delta_L^2-17\DL+4)\over 4838400(\DL-2)(\DL-3)},\cr 
    a_{333} &= {\DL(\DL-4)(\Delta_L^3-16\Delta_L^2+51\DL+24)\over 10368000(\DL-2)(\DL-3)}.
}}
We can also consider higher values of $p$ and obtain an overconstrained system of equations, 
whose solution is still \ResultTripleFourD.
Inserting  \ResultTripleFourD\ into \AnzThird, 
we obtain  the contribution from minimal-twist triple-stress tensor operators to the heavy-heavy-light-light correlator in the lightcone limit.

Note that for $\DL\to\infty$, the correlator is determined by the exponentiation of the stress tensor discussed e.g.\ in \KulaxiziTKD, i.e.\
\eqn\cub{
  \GG^{(3)}(z,\zbar)\lzbarone
  {(1-\zbar)^3\over[(1-z)(1-\zbar)]^{\DL}}\,{1\over 3!}\,\,\left({\DL\over 120}(1-z)^3{}_2F_1(3,3;6;1-z)\right)^3 +\cdots\,,
}
which one indeed obtains by taking $\DL\to\infty$ of \AnzThird\ with \ResultTripleFourD. Here ellipses denote terms subleading in $\Delta_{L}$. 

By analytically continuing $z\to e^{-2\pi i}z$ and sending $z\to 1$, one can access  the large impact parameter regime of the Regge limit. To do this we use the following property of the hypergeometric function 
(see e.g. \HartmanLFA):
\eqn\analyCont{
  {}_2F_1(a,a,2a,1-ze^{-2\pi i}) = {}_2F_1(a,a,2a,1-z)+2\pi i {\Gamma(2a)\over \Gamma(a)^2}{}_2F_1(a,a,1,z).
}
Using \analyCont\ the leading term from \AnzThird\ with the coefficients \ResultTripleFourD\ in the limit $1-\zbar \ll 1-z\ll 1$ is given by 
\eqn\leadRegge{\eqalign{
  &\GG^{(3)}(z,\zbar)\lzbaronezone{1\over [(1-z)(1-\zbar)]^{\DL}}\times \cr 
  &\left(-{9i\pi^3\DL(\DL+1)(\DL+2)(\DL+3)(\DL+4)\over 2 (\DL-2)(\DL-3)}\left({1-\zbar\over (1-z)^2}\right)^3\right).
}}  
This agrees with the holographic calculation in a shockwave background at $\OO(\mu^3)$ given by Eq.\ (45) in \FitzpatrickEFK\  based on  techniques developed in \refs{\CornalbaXK\CornalbaXM\CornalbaZB\CornalbaQF-\CostaCB}.


\subsec{Exponentiation of leading-twist multi-stress tensors}
\noindent In $d=2$ the heavy-heavy-light-light correlator is determined by the heavy-heavy-light-light Virasoro vacuum block. This block contains the exchange of any number of stress tensors and derivatives thereof in the T-channel \refs{\FitzpatrickVUA,\FitzpatrickZHA,\CollierEXN}, and therefore all multi-stress tensor contributions. This block, together with the disconnected part, exponentiates as 
\eqn\hhlldtwo{
	\langle \OO_H(\infty) \OL(1) \OL(z) \OO_H(0) \rangle = e^{\DL \FF(z)},
}
for a known function $\FF(z)$ independent of $\DL$. It is interesting to ask if something similar happens for the contribution of the  minimal-twist multi-stress tensors in the lightcone limit of the correlator in higher dimensions. By this we mean whether the stress tensor sector of the correlator can be written as
\eqn\expCorr{
  \GG(z,\zbar) \lzbarone {1\over[(1-z)(1-\zbar)]^{\DL}}e^{\DL \FF(\mu;z,\zbar)},
}
for some function $\FF(\mu;z,\zbar)$ which is a rational function of $\DL$ and remains $\OO(1)$ as $\DL\to\infty$. 

The $\zbar$ dependence implies the following form of  $\FF(\mu;z,\zbar)$:
\eqn\anzfFunc{
  \FF(\mu;z,\zbar) = \mu(1-\zbar)\FF^{(1)}(z)+\mu^2(1-\zbar)^2\FF^{(2)}(z)+\mu^3(1-\zbar)^3\FF^{(3)}(z)+\OO(\mu^4).
}
At leading order we observe  $\FF^{(1)}(z)={1\over 120}f_3(z)$, which is just the stress tensor contribution. 
At second order we find:
\eqn\ftwoFunc{\eqalign{
  \FF^{(2)}(z) = {(12-5\DL)f_3(z)^2+{15\over 7}(\DL-8)f_2(z)f_4(z)+{40\over 7}(\DL+1)f_1(z)f_5(z)\over 28800(\DL-2)}
}.}
Note that $\FF^{(2)}(z)$ is independent of $\DL$ in the limit  $\DL\to\infty$.

To find $\FF^{(3)}(z)$ we parametrise it as
\eqn\ffthreeanz{\eqalign{
	\FF^{(3)}(z)= \Big(&b_{117}f_{1}^2(z)f_7(z)+b_{126}f_1(z)f_2(z)f_6(z)+b_{135}f_1(z)f_3(z)f_5(z)\cr
  &+b_{225}f_2^2(z)f_5(z)+b_{234}f_2(z)f_3(z)f_4(z)+b_{333}f_3^3(z)\Big).
}}
It is clear that for terms which do not contain a factor of $f_3(z)$, the coefficients $b_{ijk}$ should satisfy $b_{ijk}=a_{ijk}/\DL$. This is not true for terms which contain a factor of $f_3$. 
Inserting $\FF^{(1)}$, $\FF^{(2)}$ and Eq. \ffthreeanz\ in \expCorr, expanding in $\mu$
and matching with \AnzThird\ yields
\eqn\fthreefunction{\eqalign{
  b_{117}&= {a_{117}\over \DL},\cr
  b_{126}&= {a_{126}\over\DL},\cr
  b_{225}&= {a_{225}\over \DL},\cr
  b_{135}&= -{11\Delta_L^2-19\DL-18\over 1209600(\DL-2)(\DL-3)},\cr 
  b_{234}&= {(\DL-2)(\DL+2)\over 1209600(\DL-2)(\DL-3)},\cr
  b_{333}&= {7\Delta_L^2-18\DL-24\over 2592000(\DL-2)(\DL-3)}.
}}
From \ftwoFunc\ and \fthreefunction, one finds that the correlator exponentiates to $\OO(\mu^3)$ in the sense described above, i.e. $\FF(\mu;z,\zbar)$ is a rational function of $\DL$ of $\OO(1)$ as $\DL\to\infty$. 

To leading order in $\DL$, exponentiation for large $\DL$ is a prediction of the AdS/CFT correspondence. The two-point function of the operator $\OL$ in the state created by the heavy operator $\OH$ is given in terms of the exponential of the (regularized) geodesic distance between the boundary points in the dual bulk geometry. For details on this, see e.g.\ \KulaxiziTKD.

\subsec{OPE coefficients of triple-stress tensors}
\noindent In this section we describe how to decompose the correlator \AnzThird\ into an infinite sum of minimal-twist triple-stress tensor operators.
In order to do this we use the following multiplication formula for hypergeometric functions \KulaxiziTKD:
\eqn\prodidentity{
  {}_2F_1(a,a;2a;w){}_2F_1(b,b;2b;w) = \sum_{m=0}^\infty p[a,b,m]w^{2m}{}_2F_1[a+b+2m,a+b+2m,2a+2b+4m,w],
}
where 
\eqn\co{\eqalign{
  &p[a,b,m]\cr
  =& {2^{-4m}\Gamma(a+{1\over2})\Gamma(b+{1\over2})\Gamma(m+{1\over2})\Gamma(a+m)\Gamma(b+m)\Gamma(a+b+m-{1\over 2})\Gamma(a+b+2m) \over  \sqrt{\pi}\Gamma(a)\Gamma(b)\Gamma(m+1)\Gamma(a+m+{1\over 2})\Gamma(b+m+{1\over2})\Gamma(a+b+m)\Gamma(a+b+2m-{1\over 2})}.
}}
It is useful to note that by using \prodidentity\ we can write a similar formula for the  functions  $f_{a}$ defined in \defFfunc :
\eqn\productf{f_{a}(z)f_{b}(z)=\sum_{m=0}^{\infty}p[a,b,m]f_{a+b+2m}(z),
}
\noindent where $p[a,b,m]$ is defined in \co. It is now clear that the correlator \AnzThird\ can be written as a double sum over functions $f_{9+2(n+m)}$. We can thus write the stress tensor sector of the correlator in the lightcone limit at $\OO(\mu^3)$ as
\eqn\sch{\GG^{(3)}(z,\zbar)\,\,\lzbarone \,\,{(1-\zbar)^3\over[(1-z)(1-\zbar)]^{\DL}}\sum_{n,m=0}^{\infty}c[m,n]f_{9+2(n+m)}(z),
} 
with 
\eqn\c{\eqalign{c[m,n]= & \big( a_{333} p[3, 3, m] p[3, 6 + 2 m, n] + a_{117} p[1, 7, m] p[1, 8 + 2 m, n] \cr 
&+ a_{126} p[2, 6, m] p[1, 8 + 2 m, n] + a_{135} p[3, 5, m] p[1, 8 + 2 m, n] \cr
&+ a_{225} p[2, 5, m] p[2, 7 + 2 m, n] + a_{234} p[3, 4, m] p[2, 7 + 2 m, n]\big),
}}
where coefficients $a_{ijk}$ are fixed in \ResultTripleFourD .

Comparing \sch\ with \LCBehaviour\ we see that the contribution at $\OO(\mu^3)$ comes from operators of the schematic form 
$:T_{\alpha \beta}T_{\gamma\delta}\partial_{\rho_{1}}\ldots\partial_{\rho_{2l}}T_{\mu \nu}:$. These operators have ${\tau \over 2}+s = 9+2l$, where $s$ is total spin $s=6+2l$. The corresponding OPE coefficients of such operators will be a sum of all contributions in \sch\ for which $n+m=l$.

Now, one can write OPE coefficients of operators of type $:T_{\alpha \beta}T_{\gamma\delta}\partial_{\rho_{1}}\ldots\partial_{\rho_{2l}}T_{\mu \nu}:$ as
\eqn\opeFourD{P^{(HH,LL);(3)}_{6,6+2l}=\sum_{n=0}^{l}c[l-n,n].
}
\noindent Let us write a few of the coefficients explicitly here:
\eqn\exOpeCoeffFourD{\eqalign{
    \mu^3P^{(HH,LL);(3)}_{6,6} &= \mu^3{\DL(3024+\DL(7500+\DL(7310+143\DL(25+7\DL))))\over 10378368000(\DL-2)(\DL-3)},\cr
    \mu^3P^{(HH,LL);(3)}_{6,8} &= \mu^3{\DL(2688+\DL(7148+\DL(9029+13\DL(464+231\DL))))\over 613476864000 (\DL-3) (\DL-2)},\cr 
    \mu^3P^{(HH,LL);(3)}_{6,10} &= \mu^3{\DL (888 + \DL (2216 + \DL (3742 + 17 \DL (181 + 143 \DL))))\over 9468531072000 (\DL-3) (\DL-2)}.
}}
\noindent  We further find that $P^{(HH,LL);(3)}_{6,6}$ and $P^{(HH,LL);(3)}_{6,8}$ agree with the expression obtained holographically in \FitzpatrickZQZ .

\newsec{Minimal-twist double-stress tensors in six dimensions}
\noindent In this section we derive the contribution of minimal-twist double-stress tensors to the heavy-heavy-light-light correlator in the lightcone limit in $d=6$. The method is  analogous to the four-dimensional case described in Section 3. 

From \genan\ we make the following ansatz for the stress tensor sector in the lightcone limit:
\eqn\anzSixD{\eqalign{
  	\GG^{(2)}(z,\zbar)&\lzbarone \quad {(1-\zbar)^4\over [(1-z)(1-\zbar)]^{\DL}}\,\,\times\cr
  	&\Big(a_{17}f_1(z)f_7(z)+a_{26}f_2(z)f_6(z)+a_{35}f_3(z)f_5(z)+a_{44}f_4^2(z)\Big).
}}   

The S-channel conformal blocks in six dimensions in the limit $\DH\to\infty$ are given by \SblockssSixD. In the lightcone limit $\zbar\to 1$ operators with $l\gg 1$ dominate and   the blocks can be approximated by 
\eqn\SblockSixDSimpl{
  (z\zbar)^{-{1\over 2}(\DH+\DL)}g_{\DH+\DL+2n+\gamma,l}^{(\DHL,-\DHL)}(z,\zbar) \simeq  {\zbar^l z^n p(\log z,\gamma)\over (1-z)^2},
}
with $p$ given by \pfunc. Replacing the sum in \sCh\ with an integral and inserting \SblockSixDSimpl\ we have
\eqn\sChAppxIntegral{
	\GG^{(2)}(z,\zbar)\lzbarone \sum_{n=0}^\infty{z^n \over (1-z)^2}\int_0^\infty dl P^{(HL,HL)}_{n,l}\zbar^l  p(\log z,\gamma).
}
As in $d=4$ one finds that terms proportional to $\log^i z$ with $i=2,3,\ldots,k$ at $\OO(\mu^k)$, are determined by the OPE data at $\OO(\mu^{k-1})$. 

At $\OO(\mu)$ we can use the known contribution from the stress tensor exchange \StressTensorCont\ to derive the anomalous dimensions $\gamma^{(1)}_n$ and the OPE coefficients $P^{(1)}_n$ just as it was done in four dimensions. 
This is  done by matching \sChAppxIntegral\ order by order in the small $z$ expansion.
 Using \Spindep\ one can  integrate  over spin. E.g.\ for $n=0,1,2,3$:
\eqn\anomdimfirstorderSixD{\eqalign{
  \gamma^{(1)}_0 &= -{(\DL-2)(\DL-1)\DL\over 2},\cr
  \gamma^{(1)}_1 &= -{(\DL-1)\DL(\DL+10)\over 2},\cr
  \gamma^{(1)}_2 &= -{\DL(\DL+2)(\DL+19)\over 2},\cr
  \gamma^{(1)}_3 &= -{(\DL+4)(\DL(\DL+29)+30)\over 2},\cr
}}
These anomalous dimensions  agree with eq.\ (6.10) in \KulaxiziDXO.
Similarly, we  obtain the following OPE coefficients:
\eqn\OPEFirstOrderSixD{\eqalign{
  P^{(1)}_0 &= -{11(\DL-2)(\DL-1)\DL\over 12},\cr 
  P^{(1)}_1 &= -{(\DL-1)\DL(11\DL+38)\over 12},\cr 
  P^{(1)}_2 &= -{\DL(22+\DL(87+11\DL))\over 12},\cr 
  P^{(1)}_3 &= -{\DL(202+\DL(147+11\DL))\over 12}.\cr
}}
It is  straightforward to continue to higher values of $n$.

Plugging \anomdimfirstorderSixD\ into \sChAppxIntegral\ in the limit $1-\zbar\ll z\ll 1$ one finds the following contribution to the  terms proportional to ${z^p\log^2 z\over (1-\zbar)^{\DL-4}}$ at $\OO(\mu^2)$:
\eqn\ResTwoD{\eqalign{
  p=0:&{\Delta_L^2(\DL-1)(\DL-2)\over 32(\DL-3)(\DL-4)},\cr
  p=1: &{\Delta_L^2(\DL-1)(\DL+6)(\DL+16)\over 32(\DL-3)(\DL-4)}\cr
  p=2: &{\Delta_L^2(\Delta_L^4+46\Delta_L^3+599\Delta_L^2+1898\Delta_L+1056)\over 64(\DL-3)(\DL-4)},\cr
  p=3:&{\Delta_L^7+72\Delta_L^6+1651\Delta_L^5+13344\Delta_L^4+40180\Delta_L^3+41952\DL^2+14400\DL\over 192(\DL-3)(\DL-4)}\,\,.
}}
It is now straightforward to expand the ansatz \anzSixD\ in the limit $1-\zbar\ll z\ll 1$ , collect terms that behave as $z^p\log^2z$ and compare them to  the S-channel \ResTwoD. This determines the coefficients: 
\eqn\a{\eqalign{
  a_{17} &= {\DL(\DL+1)(\DL+2)\over 64064(\DL-3)(\DL-4)},\cr
  a_{26} &= {\DL(-18+(-12+\DL)\DL)\over 133056 (\DL-3)(\DL-4)},\cr
  a_{35} &= {\DL(\DL-6)(\DL-15)\over 302400(\DL-3)(\DL-4)},\cr
  a_{44} &= {\DL(\DL-5)(\DL-6)\over 627200(\DL-3)}.
}}
One can consider higher values of $p$; eq. \a\ is still the solution of the corresponding overconstrained system.

The double-stress tensor contribution to the correlator in the lightcone limit $\zbar\to 1$ is therefore given by 
\eqn\ResultCorrSixD{\eqalign{
  \GG^{(2)}&(z,\zbar)\lzbarone{(1-\zbar)^{4}\over [(1-z)(1-\zbar)]^{\DL}}{\DL\over (\DL-3)(\DL-4)}\left({1\over 627200}\right)\cr
  &\times\Big\{(\DL-4)(\DL-5)(\DL-6)f_{4}^2(z)+{56(\DL-6)(\DL-15)\over27}f_3(z)f_5(z)\cr
  &+{1400(\DL(\DL-12)-18)\over 297}f_2(z)f_6(z)+{1400(\DL+1)(\DL+2)\over 143}f_1(z)f_7(z)\Big\}.
}}

Using  \ResultCorrSixD\ one can deduce the second order OPE data in the S-channel. 
The anomalous dimensions at this order can then be compared 
to the holographic calculations in \KulaxiziDXO\ to reveal perfect agreement.

\subsec{Exponentiation of minimal-twist multi-stress tensors in six dimensions}
\noindent It is interesting to study whether  the stress tensor sector of the correlator exponentiates in the lightcone limit 
\eqn\correxpSixd{
  \GG(z,\zbar)\lzbarone{1\over [(1-z)(1-\zbar)]^\DL}e^{\DL \FF(\mu;z,\zbar)},
}
with $\FF(\mu;z,\zbar)$ a rational function of $\DL$ that is of $\OO(1)$ as $\DL\to\infty$.
In the lightcone limit $\FF(\mu;z,\zbar)$ admits an expansion
\eqn\FSixD{
  \FF(\mu;z,\zbar)=\mu(1-\zbar)^2\FF^{(1)}(z)+\mu^2(1-\zbar)^4\FF^{(2)}(z)+\OO(\mu^3). 
}
At $\OO(\mu)$  one finds $\FF^{(1)}(z)={\Gamma({6\over 2}+1)^2\over 4\Gamma(6+2)}f_4(z)$ from the stress tensor contribution. 
Using  \ResultCorrSixD\ we find
\eqn\FSixDtwo{   \FF^{(2)}(z)=b_{17}f_1(z)f_7(z)+b_{26}f_2(z)f_6(z)+b_{35}f_3(z)f_5(z)+b_{44}f_4^2(z) }
with
\eqn\bcoeffSixD{\eqalign{
  b_{17} &= {a_{17}\over \DL},\cr
  b_{26} &= {a_{26}\over \DL},\cr
  b_{35} &= {a_{35}\over \DL},\cr
  b_{44} &= -{4\Delta_L^2-31\DL+60\over 313600(\DL-3)(\DL-4)}.
}}  
From \bcoeffSixD\ we indeed see that the stress tensor sector of the correlator exponentiates  
at least to $\OO(\mu^2)$ in $d=6$.

\subsec{OPE coefficients of minimal-twist double-stress tensors}
\noindent In this Section we decompose the stress tensor sector of the correlator \anzSixD\ into a sum over minimal-twist double-stress tensors. 
The discussion follows that of Section 3.5.

Applying \productf\ to \ResultCorrSixD, we find that $a+b+2l=8+2l$ which is ${\tau\over 2}+s+2l$, with $\tau=8$ and $s=4$ being the twist and spin of the simplest minimal-twist double-stress tensor operator $:T_{\mu\nu}T_{\rho\lambda}:$. Non-zero value of $l$ thus gives the contribution from operators of higher spin of the form $:T_{\mu\nu}\partial_{\rho_{1}}\ldots\partial_{\rho_{2l}}T_{\delta\lambda}:$, where no indices are contracted and only even spin operators contribute to the OPE between identical scalars.   

It is now straightforward to write down the OPE coefficients for minimal-twist double-stress tensors in six dimensions. E.g.\ one finds for the lowest-spin operators the following OPE coefficients:
\eqn\exOpeCoeff{\eqalign{
  \mu^2P^{(HH,LL);(2)}_{8,4} &= \mu^2{\DL(600+\DL(1394+\DL(677+429\DL)))\over 269068800(\DL-3)(\DL-4)},\cr 
  \mu^2P^{(HH,LL);(2)}_{8,6} &= \mu^2{\DL(30+\DL(187+\DL(-120+143\DL)))\over 3430627200(\DL-3)(\DL-4)},\cr
  \mu^2P^{(HH,LL);(2)}_{8,8} &= \mu^2{\DL(60+\DL(1382+\DL(-1857+1105\DL)))\over 657033721344(\DL-3)(\DL-4)}.
}}
For general spin we have ($s=4+2l$)
\eqn\opeCoeffSixDTT{
   P^{(HH,LL)}_{8,s} = \mu^2{c\DL\over (\DL-3)(\DL-4)}(a_3\Delta_L^3+a_2\Delta_L^2+a_1\DL+a_0)
}
where 
\eqn\cods{\eqalign{
  c &= {2^{-9-2s}\sqrt{\pi}s(s+2)\Gamma(s-1)\over (s-3)(s+4)(s+6)(s+8)(s+10)\Gamma(s+{7\over 2})},\cr
  a_3&= (s-2)s(s+2)(s+5)(s+7)(s+9),\cr
  a_2&=-3(2880+s(s+7)(-276+s(s+7)(-56+s(s+7)))),\cr
  a_1&= 2(25920+s(s+7)(3276+s(s+7)(-80+s(s+7)))),\cr
  a_0&= 675\times 2^7.
}}

\newsec{Discussion}
\noindent In this paper we consider the minimal-twist multi-stress tensor contributions to the heavy-heavy-light-light correlator of scalars in large $C_T$ CFTs in even spacetime dimensions. 
We provide strong evidence for the conjecture that all such contributions are described by the ansatz \genan\ and determine the coefficients by 
performing  a bootstrap procedure.
In practice this is completed for twist-four double-stress tensors and twist-six triple-stress tensors in four dimensions as well as twist-eight double-stress tensors in six dimensions. 
 In principle it is straightforward to use our technology to determine the coefficients $a_{i_1 ... i_k} $ to arbitrarily high order in $\mu$;
 this must be  related to the universality of the minimal-twist OPE coefficients.

In two dimensions the heavy-heavy-light-light Virasoro vacuum block exponentiates [see eq. \virvac ], with $\FF(\mu;z)$ independent of $\DL$. 
In higher dimensions we observe a similar exponentiation with $\FF(\mu;z,\zbar)$ a rational function of $\DL$ that remains $\OO(1)$ as $\DL\to\infty$. 
 It would be interesting to see whether it is possible to write down a closed-form recursion formula for $\FF(\mu;z,\zbar)$. 
 Solving such a recursion formula would give a higher-dimensional analogue of the two-dimensional Virasoro vacuum block.

An immediate technical question concerns CFTs in odd spacetime dimensions.
We could not immediately generalize our results in this context -- the ansatz in eq.\ \genan\ fails in odd dimensions. However, the heavy-light conformal blocks are known \KarlssonQFI, so a similar approach should be feasible. 

It would be interesting to study the regime of applicability of our results. 
We have not used holography; our main assumption 
is the ansatz \gkhd, known to be true for holographic CFTs to $\OO(\mu^2)$ in $d=4$ \KulaxiziTKD. 
Yet, our general expressions for the OPE coefficients agree with the OPE coefficients computed in 
some holographic examples  
\FitzpatrickZQZ.
What happens once one goes beyond holographic CFTs - 
will our ansatz need to be modified by the inclusion of terms suppressed by the gap or the central charge? We leave these questions for subsequent investigations.

Another interesting direction concerns the  study of the bulk scattering phase-shift in the presence of a black hole background.
In the context of higher dimensional CFTs, this problem was first considered in \KulaxiziDXO\ where the gravitational expression was given to all orders in $\mu$ and the CFT computation was performed to $\OO(\mu)$. 
Subsequently, $\OO(\mu^2)$ was discussed in \KarlssonQFI.
In \FitzpatrickEFK\ the $\OO(\mu)$ contribution was exponentiated to yield the scattering phase shift in the presence of a shock-wave geometry. 
A CFT computation  of the phase shift to all orders in $\mu$ is still lacking. 
This would in principle involve understanding Regge theory beyond the leading order. 
It will be interesting to see whether the results of this article could be helpful in this regard.

\bigskip
\bigskip

\noindent {\bf Acknowledgments}: 
We thank G-S Ng, E. Perlmutter, K. Sen and D. Simmons-Duffin for useful discussions. The work of R.K. and A.P. is supported in part by an Irish Research Council Laureate Award. The work of P.T. is supported in part by an Ussher Fellowship Award. M.K. and A.P. thank the  Aspen Center for Physics, where part of this work was completed, for hospitality. Work at ACP is supported by the National Science Foundation under Grant No. NSF PHY-1607611.

\listrefs

\bye